\documentclass[amsmath,pra,amssymb,aps,floatfix,notitlepage,twocolumn]{revtex4-2}

\usepackage{graphicx}
\usepackage{xcolor}
\usepackage{booktabs}

\usepackage{braket}
\usepackage{float} 
\usepackage{setspace}
\usepackage{mathrsfs}
\usepackage{hyperref}

\newcommand{\deltaf}{\delta\!f}

\begin{document}

\title{Resistivity of the two-dimensional Bose-Hubbard model at weak coupling}
\author{Eduardo O. Rizzatti}
\email{eo279@cornell.edu}
\affiliation{Laboratory of Atomic and Solid State Physics, Cornell University, Ithaca, New York 14853, USA}
\author{Erich J. Mueller}
\email{em256@cornell.edu}
\affiliation{Laboratory of Atomic and Solid State Physics, Cornell University, Ithaca, New York 14853, USA}
\date{\today}

\begin{abstract}
We calculate the weak-coupling resistivity of the two-dimensional Bose Hubbard model, comparing with the more familiar fermionic case.  At high temperature  the resistivity is linear in $T$, while in the low temperature normal state it is exponentially suppressed.  We explore the density dependence and calculate the momentum relaxation rate.
\end{abstract}
\maketitle

\section{Introduction}\label{sec:intro}
Transport is the most fundamental probe in solid state physics.  Recently, new techniques have been developed to study the transport properties of quantum degenerate atomic gases \cite{brown_bad_2019,anderson_conductivity_2019,xu_bad-metal_2019}, giving us the ability to use ultracold atoms to expand our understanding.  For example, many strongly correlated materials display unexplained temperature dependence of their resistivity \cite{grigera_magnetic_2001,hussey_phenomenology_2008,bruin_similarity_2013,legros_universal_2019}, and  cold atom experiments may reveal the underlying mechanisms.   Cold atom experiments are also exploring the rich behavior of transport in inhomogeneous settings such as quantum point contacts \cite{van_houten_quantum_1996}.  Beyond these solid state inspired studies, we can use atomic gases to explore completely novel transport regimes.  Here we calculate the resistivity of a 2D Bose gas in an optical lattice.  As is familiar from other systems with bounded spectra \cite{mukerjee_statistical_2006,lindner_conductivity_2010,perepelitsky_transport_2016,huang_strange_2019,patel_many-body_2022} we find that the high temperature resistivity scales linearly in temperature, while the low temperature behavior is more complex.  The same techniques which were used to measure the resistivity of a 2D Fermi gas could be applied here to experimentally verify these results.

In most materials, resistivity is dominated by either electron-phonon or electron-impurity scattering.  Ultracold atoms have no phonons or impurities, so the only mechanism for momentum relaxation is the equivalent of electron-electron scattering~\cite{rosch_optical_2006}.  Momentum is conserved in these processes, modulo a reciprocal lattice vector.  Resistivity requires momentum relaxation, and is thus related to the rate of umklapp scattering~\cite{peierls_quantum_1996}, where the momentum of the particles change by such a reciprocal lattice vector.  We calculate this rate within the Born approximation, which works well at weak coupling, extracting the resistivity.

Resistivity is a quantity which is almost exclusively discussed in the context of fermions.  Bosons, such as light, typically move ballistically.  Moreover, there is no obvious optical equivalent of an electric field, making it challenging to even envision an optical transport experiment.  Other common bosonic systems, such as phonons or magnons suffer these same challenges.  Cold atoms, however, provide the perfect platform to extend transport measurements to bosons.   All of the techniques that have been developed to probe transport in fermionic gasses can be used on bosonic atoms.  Furthermore, the same mechanism which lead to the ohmic behavior of fermions  also apply to bosons.    Due to the lack of a Fermi surface, however, much of the phenomenology differs.  Additionally, at very low temperatures a Bose gas becomes superfluid, with a vanishing resistivity.  Our calculations only apply to the normal state.

At infinite temperature a lattice gas will have infinite resistivity \cite{mukerjee_statistical_2006,lindner_conductivity_2010,kiely_transport_2021}.  All states are equally likely, and a uniform force does not change the momentum distribution or lead to currents.  The leading corrections scale with $1/T$, giving a $T$-linear high temperature resistivity.  This argument is equally applicable to bosons and fermions.

For fermions~\cite{kiely_transport_2021} at low temperature, all properties are dominated by excitations around the Fermi surface.  Counting arguments then give that the umklapp contribution to resistivity generically scales as $T^2$.  Bosons are different.  The low-temperature (but non-condensed) Bose distribution is peaked at small momentum.  There are no allowed umklapp processes for these small momentum particles, so the resistivity is exponentially small.

At very low temperatures a 2D interacting Bose gas will undergo a Berezinskii–Kosterlitz-Thouless transition to a superfluid state \cite{berezinsky_destruction_1971,berezinsky_destruction_1972,kosterlitz_ordering_1973,fisher_dilute_1988, al_khawaja_low_2002}, in which our calculation does not apply. 
The transition occurs at a temperature of order $T_{BKT}\sim n t/\ln (t/nU)$  where $n$ is the average number of particles per site, $t$ is the hopping matrix element and $U$ is the onsite interaction strength.  Typical densities are of order $0.1<n<2$. 

We organize our paper as follows. In Sec.~\ref{sec:model} we describe the 2D Bose-Hubbard model and our variational approach
to solve the Boltzmann equation. In Sec.~\ref{sec:results}
we present our results for the resistivity and its temperature dependence.  In Sec.~\ref{sec:conclusions} we summarize and present some conclusions.

\section{The Model}\label{sec:model}

We consider spinless bosonic particles on a square lattice, whose dynamics are governed by the Bose-Hubbard Hamiltonian~\cite{fisher_boson_1989}
\begin{eqnarray}\label{hamiltonian}
H &=& -t\sum_{\langle i, j \rangle} b^{\dagger}_ib^{\mathstrut}_j + \frac{U}{2}\sum_{i} b^{\dagger}_ib^{\dagger}_ib^{\mathstrut}_ib^{\mathstrut}_i  \\ \nonumber
&=& \sum_{\mathbf{k}} \epsilon^{\mathstrut}_\mathbf{k} b_\mathbf{k}^\dagger b^{\mathstrut}_{\mathbf{k}} + \frac{U}{2N_s}\sum_{\mathbf{k} \mathbf{k}' \mathbf{q}} b^{\dagger}_{\mathbf{k}} b^{\dagger}_{\mathbf{k}'} b^{\mathstrut}_{\mathbf{k'-q}} b^{\mathstrut}_{\mathbf{k+q}} \;.
\end{eqnarray}
where $b^{\dagger}$/$b$ are the creation/annihilation operators, $t$ is the hopping amplitude and $U$ accounts for the local interaction. The single-particle dispersion relation is
\begin{equation}\label{energy_dispersion}
\epsilon_{\mathbf{k}} = -2t[\cos(k_xa)+\cos(k_ya)] \;,
\end{equation}
where $a$ is the lattice spacing, and the momentum space sums are taken over $k=(k_x,k_y)$ with $k_x,k_y=2\pi n/L$, where $N_s$  is the number of lattice sites and $V=N_sa^2=L^2$ is the volume (area).  

We will use a quantum kinetic approach~\cite{uehling_transport_1933,kadanoff_quantum_2018}, describing the system in terms of the distribution function $f_\mathbf{k}=f_\mathbf{k}(\mathbf{r},t)$, which measures how many particles have momentum $\mathbf{k}$ at position $\mathbf{r}$ and time $t$.  We use the same symbol for the hopping matrix element and time, but the meaning should be clear from context.

The distribution function obeys a Boltzmann equation 
\begin{equation}\label{bte}
\frac{\partial f_\mathbf{k}}{\partial t} +  \mathbf{v}_{\mathbf{k}} \cdot \nabla_{\mathbf{r}} f_\mathbf{k} + \frac{e \mathbf{E}}{\hbar} \cdot \nabla_{\mathbf{k}} f_\mathbf{k} = I_{\mathbf{k}}
\end{equation}
Here $\mathbf{v}_{\mathbf{k}} = \frac{1}{\hbar}\nabla_{\mathbf{k}} \epsilon_{\mathbf{k}}$.   The electric field $\bf E$ is small, and we have taken the particles to have charge $e$. In the cold atom context the atoms are neutral and this designation is formal, since the only physical quantity is the gradient of the potential energy, $\nabla \mathcal{V}=-e E$.
The collision integral expresses the 
change in momentum occupation due to scattering,
\begin{equation}
I_{\mathbf{k}} = \sum
\left[ \Omega_{\rm in}-\Omega_{\rm out} \right] \mathcal{W}_{\mathbf{k} \mathbf{k'}}^{\mathbf{k''} \mathbf{k'''}} 
\label{collisional_integral}
\end{equation}
where the sum is over $\mathbf{k^\prime,k^{\prime\prime},k^{\prime\prime\prime}}$ in the first Brillioun zone ($-\pi/a<k_x,k_y<\pi/a$). 
Quantum statistics are encoded in the factors $\Omega_{\rm in}=f_{\mathbf{k'''}}f_{\mathbf{k''}}(1+f_{\mathbf{k'}})(1+f_{\mathbf{k}})$ and
$\Omega_{\rm out}=f_{\mathbf{k}}f_{\mathbf{k'}}(1+f_{\mathbf{k''}})(1+f_{\mathbf{k'''}})$ .
Using Fermi's golden rule, the 
transition rate $\mathcal{W}_{\mathbf{k} \mathbf{k'}}^{\mathbf{k''} \mathbf{k'''}} =
\frac{2\pi}{\hbar}|\bra{\mathbf{k''}\mathbf{k'''}}\mathcal{H}\ket{\mathbf{k}\mathbf{k'}}|^2\delta (\epsilon_{\mathbf{k}}+\epsilon_{\mathbf{k'}}-\epsilon_{\mathbf{k''}}-\epsilon_{\mathbf{k'''}})$ 
can be evaluated as
\begin{equation}\label{transition_rate}
\mathcal{W}_{\mathbf{k} \mathbf{k'}}^{\mathbf{k''} \mathbf{k'''}} =
\frac{U^2}{4N_s^2} \bar\delta_{\mathbf{k}+\mathbf{k}',\mathbf{k}''+\mathbf{k}'''} \delta (\epsilon_{\mathbf{k}}+\epsilon_{\mathbf{k'}}-\epsilon_{\mathbf{k''}}-\epsilon_{\mathbf{k'''}}) 
\end{equation}
where $\mathcal{H}= \frac{U}{2}\sum_{i} b^{\dagger}_ib^{\dagger}_ib^{\mathstrut}_ib^{\mathstrut}_i$ is the interaction Hamiltonian. Here $\bar \delta_{\mathbf{pq}}$ is the periodic version of the Kronecker delta function, which is zero unless $\mathbf{p-q}$ is a reciprocal lattice vector.

We linearize the Boltzmann equation by considering a small perturbation $\deltaf_{\mathbf{k}} = -\Phi_{\mathbf{k}}({\partial f_\mathbf{k}^{0}}/{\partial\epsilon_{\mathbf{k}}})$ from the local equilibrium distribution function, $f_{\mathbf{k}}^{0} = (e^{\beta(\epsilon_{\mathbf{k}}-\mu)}-1)^{-1}$, writing
\begin{eqnarray}
f_{\mathbf{k}} = f_{\mathbf{k}}^{0} 
- \Phi_{k}\frac{\partial f_\mathbf{k}^{0}}{\partial\epsilon_{\mathbf{k}}} \;,
\label{distribution_function}
\end{eqnarray}
where the temperature $k_BT=\beta^{-1}$ and the chemical potential $\mu$ are uniform and time independent, implying $\nabla_{\mathbf{r}}f^{0}_{\mathbf{k}}\approx 0$. The function $\Phi_\mathbf{k}$ is small and can be interpreted as a generalized current, generating a deviation from the equilibrium distribution. In a steady state configuration $\partial_{t} f_{\mathbf{k}}=0$ , and the linearized Boltzmann equation becomes $e (\partial f_\mathbf{k}^0/\partial\epsilon_\mathbf{k})\mathbf{ v_k\cdot E}=I_\mathbf{k}$ with  
\begin{eqnarray}\label{collision_integral_p}
I_{\mathbf{k}}&=& -\beta \sum
\left[ \Phi_{\mathbf{k}}+\Phi_{\mathbf{k'}}-\Phi_{\mathbf{k''}}-\Phi_{\mathbf{k'''}} \right] \mathcal{P}_{\mathbf{k} \mathbf{k'}}^{\mathbf{k''} \mathbf{k'''}} \\
\label{probability_rate}
\mathcal{P}_{\mathbf{k} \mathbf{k'}}^{\mathbf{k''} \mathbf{k'''}}&=&
f_{\mathbf{k}}^0f_{\mathbf{k'}}^0(1+f_{\mathbf{k''}}^0)(1+f_{\mathbf{k'''}}^0) \mathcal{W}_{\mathbf{k} \mathbf{k'}}^{\mathbf{k''} \mathbf{k'''}}.
\end{eqnarray}

Following the approach in \cite{ziman_general_1956,ziman_electrons_2001}, we make a variational ansatz, $\Phi_{\mathbf{k}} =  \xi \phi_{\mathbf{k}}$, where $\phi_{\mathbf{k}}=(\nabla_{\mathbf{k}} \epsilon_{\mathbf{k}})_x = 2t \sin(k_xa)$ is fixed and $\xi$ is a variational parameter.  A more sophisticated ansatz would involve a linear combination of a set of  linearly independent trial functions. If such set is complete, then the solution is exact. We 
multiply the Boltzmann equation by $-\Phi_{\mathbf{k}}/V$ and sum over $\mathbf{k}$, resulting in
\begin{equation}\label{entropy_balance}
\xi \mathbf{j}\cdot \mathbf{E} 
=\xi^2 P.
\end{equation}
where 
\begin{eqnarray}\label{current}
\mathbf{j} &=& -\frac{e}{V}\sum_{\mathbf{k}}  \phi_{\mathbf{k}}
\frac{\partial f^0_k}{\partial \epsilon_k}
\mathbf{v}_{\mathbf{k}}  \\
\label{pexp}
P&=& \frac{ \beta}{4 V} \sum
\left[ \phi_{\mathbf{k}}+\phi_{\mathbf{k'}}-\phi_{\mathbf{k''}}-\phi_{\mathbf{k'''}} \right]^2 \mathcal{P}_{\mathbf{k} \mathbf{k'}}^{\mathbf{k''} \mathbf{k'''}} 
\end{eqnarray}
Physically, these encode the charge density  $\mathbf{J}=
\xi\mathbf{j}$, and the entropy production due to collisions  $\dot{S}_{coll}=\xi^2 P/T$. 
The connection between $P$ and entropy comes from noting that the left hand side of 
Eq.~(\ref{entropy_balance}) is the power from ohmic dissipation $\mathbf{J}\cdot \mathbf{E} = -T\dot{S}_{field}$, which is converted into heat by collisions.

As argued in Appendix~\ref{appendix:variational_procedure}, the $\xi={\bf j\cdot E}/P$ which satisfied Eq.~(\ref{entropy_balance}) maximizes the entropy production, and provides our best variational estimate. The resistivity is then 
\begin{equation}
\rho=\frac{P}{j^2}.
\end{equation}
where $j=j_x$ is the $x$-component of the current, parallel to the applied electric field. We also define a relaxation time $\tau$ from the Boltzmann equation according to $-\frac{\deltaf_{\mathbf{k}}}{\tau} = I_{\mathbf{k}}$.
Multiplying both sides by $\Phi_{\mathbf{k}}/V$ and summing over $\mathbf{k}$, the momentum relaxation rate $\Gamma =\tau^{-1}$ is
\begin{equation}\label{scattering_rate}
\Gamma = \frac{ea}{\hbar}\frac{P}{j} \;.
\end{equation}

For a given temperature and density, $P$ is a 8 dimensional integral, but conservation of energy and momentum reduce it to $5$ dimensions.  The current $j$ is a 2 dimensional integral.  In Appendix~\ref{appendix:P} we explain how to efficiently calculate $P$.

\section{Results}\label{sec:results}

\begin{figure}
    \centering
    \includegraphics[scale=1.]{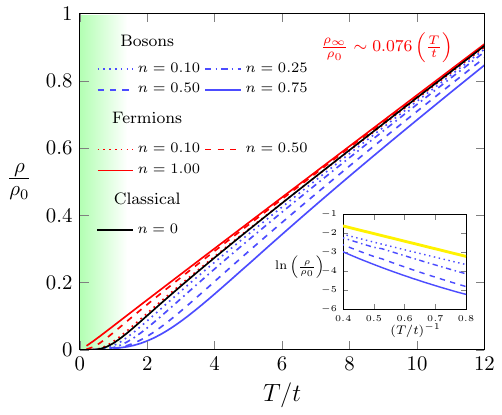}
    \caption{
Resistivity $\rho$ as a function of temperature $T$ for bosons (blue) or fermions (red) hopping on a 2D square lattice in the weak coupling limit, $t\gg U$.  Here $\rho_0=\frac{\hbar}{e^2}\left(\frac{U}{t}\right)^2$ is quadratic in the on-site interaction strength $U$. The hopping matrix element is $t$.  The black line shows the classical limit, corresponding to vanishing density.  Inset shows the bosonic data, on an inverse semilog scale.  The thick yellow line has a slope of $-4$, corresponding to $\rho\propto e^{-4t/T}$.  At sufficiently low temperature, $T_{\rm BKT}\sim nt/\ln(t/nU)$, the system will undergo a BKT transition to a superfluid, where this calculation is invalid.
The shaded region indicates typical parameter values for the transition.
    }
    \label{fig:resistivity}
\end{figure}

Our results are summarized by Fig.~\ref{fig:resistivity}.  Within the Born approximation, the resistivity scales with $U^2$, and has no other interaction dependence.  The characteristic scale is $\rho_0=\frac{\hbar}{e^2}\left(\frac{U}{t}\right)^2$.
At high temperatures the resistivity is linear in $T$, approaching $\rho_{\infty} = 0.076 \left(\frac{T}{t}\right) \rho_0$.  Such linear behavior is expected for any system with a bounded spectrum \cite{mukerjee_statistical_2006,lindner_conductivity_2010,perepelitsky_transport_2016,huang_strange_2019,patel_many-body_2022}.

The resistivity monotonically decreases with increasing density.  This feature is related to the fact that the only momentum changing collisions involve umklapp processes, $\mathbf{k,k'\to k'',k'''}$ with $\mathbf{k+k'= k''+k'''+ Q}$, where $\mathbf{Q}$ is a reciprocal lattice vector.  If the energies $\epsilon_\mathbf{k}$ and $\epsilon_{\mathbf{k}'}$ are sufficiently small, then there are no energy conserving processes of this form.  Thus low energy particles have a smaller contribution to the resistivity than those with large energy.  For bosons, increasing the density will increase the fraction of particles with low energy, resulting in a suppressed resistivity.  

Conversely, for fermions increasing the density shifts the particle distribution to higher energy.  Thus the resistivity grows with density (below half-filling). In Fig.~\ref{fig:resistivity} we also show the fermion result, as calculated in \cite{kiely_transport_2021}.  At vanishing density the bosonic and fermionic results coincide, corresponding to a Boltzmann gas.

At low temperature the resistivity is strongly suppressed, as exponentially small numbers of particles have sufficient energy to undergo umklapp processes.  Thus $\rho\sim e^{-\Delta/T}$.  The inset illustrates this behavior by plotting $\ln(\rho/\rho_0)$ as a function of temperature $1/T$.  For all densities the data is well fit to $\Delta=4 t$, corresponding to the fact that in this regime the most important collisions occur between particles near the bottom of the band (with $\epsilon=-4t$) and particles at the van-Hove singularity (with $\epsilon=0$).

\begin{figure}
    \centering
    \includegraphics[scale=1]{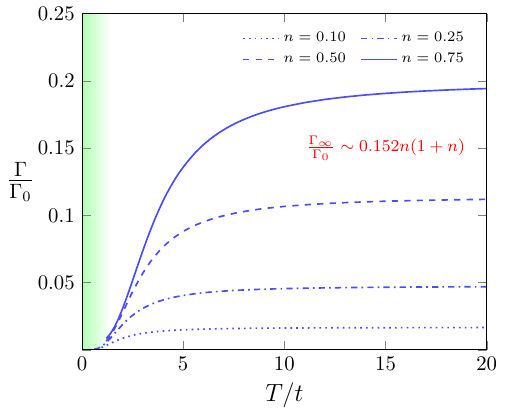}
    \caption{Momentum relaxation rate $\Gamma$ versus temperature $T$ for bosons, at several different densities $n$. At high temperatures, $\Gamma$ saturates at a value $\Gamma_{\infty} = 0.152 n(1+n)U^2/\hbar t$. For low temperatures,  the relaxation rate is exponentially suppressed.}
    \label{fig:scattering_rate}
\end{figure}

In addition to looking at the resistivity, it is useful to also explore the temperature dependence of the momentum relaxation rate, $\Gamma$. While direct measures of $\Gamma$ are often challenging in solid state physics, cold gas experiments have demonstrated the ability to measure it~\cite{barker_accurate_2023}. Figure~\ref{fig:scattering_rate} shows $\Gamma$ versus $T$ for bosons at different densities.
The relaxation rate is a monotonically increasing function of density and temperature.  More particles provide more opportunities for scattering, as well as contributing via Bose enhancement.  Higher temperature result in the occupation of larger momenta, and hence more opportunities for umklapp scattering.
At high temperatures the distribution function becomes flat and  $\Gamma$ saturates at  $\Gamma_{\infty} = 0.152 n(1+n) \Gamma_0$, with $\Gamma_0=U^2/t\hbar$.  

For low temperatures, $\Gamma$ becomes exponentially suppressed, due to the low occupation of high momentum modes.  One also notes that the relaxation rate has only weak $n$ dependence at low $T$.  This is due to the fact that at sufficiently low temperatures, $\mu\to -4 t$ for all densities.  Thus the occupation of high momentum modes becomes density independent.

\section{Summary and Conclusions}\label{sec:conclusions}

Cold atom experiments give us access to novel transport quantities, such as the resistivity of a Bose gas.  Thus we can elucidate the role of quantum statistics in transport.

Here we calculated the weak coupling resistivity of a Bose gas in a 2D optical lattice.  We find $\rho\propto T$ at high temperatures, similar to what was seen in experiments on fermionic atoms \cite{brown_bad_2019}.  This illustrates that high $T$ linear resistivity is a property of the bound spectrum and is not solely a signature of exotic strongly-correlated systems.  In this high temperature regime quantum statistics manifest in the density dependence of the resistivity.  For bosons the resistivity falls with increasing density, while the opposite occurs for fermions.

At low temperature we instead find that $\rho$ is exponentially suppressed, as opposed to the fermionic $\rho \propto T^2$.  This difference is due to the momentum distribution of the particles.  At low temperatures the bosons predominantly occupy low $k$ modes, which are incapable of umklapp scattering.  Conversely, fermions form a Fermi sea.  We also demonstrate that in the limit of vanishing density quantum statistics become irrelevant, and the fermionic and bosonic results coincide.

Experiments on bosons have largely focused on the superfluid regime.  As this study illustrates, however, there are also many interesting phenomena which can be explored in the normal state.  It would be very exciting to experimentally observe ohmic flow in a Bose gas.  One could also explore the interplay between hydrodynamic and dissipative transport, as has been studied in recent solid state experiments \cite{lucas_hydrodynamics_2018}.

\section{acknowledgements}
We thank Thomas Kiely for helpful conversations.
This material is based upon work supported by the National Science Foundation under Grant No. PHY- 2110250.  Eduardo Rizzatti thanks CNPq for the postdoc scholarship, grant 401867/2022-6.

\appendix

\section{Numerically evaluating $P$ }\label{appendix:P}

In the thermodynamic limit the discrete sums over momentum  in Eq.~(\ref{pexp}) become integrals and the periodic Kroenecker delta transform into  a sum of Dirac Delta functions according to 
\begin{eqnarray} \label{thermodynamic_limit}
\frac{1}{V}\sum_{\mathbf{k}} & \rightarrow &   \int \frac{d^2\mathbf{k}}{(2\pi)^2} \\
V\bar\delta_{\mathbf{k}+\mathbf{k'}, \mathbf{k''}+\mathbf{k'''}} & \rightarrow & \sum_Q (2\pi)^2 \delta (\mathbf{k}+\mathbf{k'}-\mathbf{k''}-\mathbf{k'''}-\mathbf{Q}) \nonumber 
\end{eqnarray}
where the reciprocal lattice vectors $Q$ have components $Q_x,Q_y$ which are integer multiples of $2\pi/a$.
The collision term $P$ is then
\begin{widetext}
\begin{eqnarray}\label{P_continuum}
P = \frac{U^2 a^4 \beta}{16\hbar(2\pi)^5} \sum_Q \int d^2\mathbf{k}  \int d^2\mathbf{k'}  \int d^2\mathbf{k''}  \int d^2\mathbf{k'''} (\phi_{\mathbf{k}}+\phi_{\mathbf{k'}}-\phi_{\mathbf{k''}}-\phi_{\mathbf{k'''}})^2  f_{\mathbf{k}}^0f_{\mathbf{k'}}^0(1+f_{\mathbf{k''}}^0)(1+f_{\mathbf{k'''}}^0) \times \nonumber \\
\times\delta(\mathbf{k}+\mathbf{k'}-\mathbf{k''}-\mathbf{k'''}-\mathbf{Q})   \delta (\epsilon_{\mathbf{k}}+\epsilon_{\mathbf{k'}}-\epsilon_{\mathbf{k''}}-\epsilon_{\mathbf{k'''}})
\end{eqnarray}
where, as already introduced, $\phi_{\mathbf{k}}=2t \sin(k_xa)$, $\epsilon_\mathbf{k}= -2t(\cos(k_xa)+\cos(k_ya))$, and $f_{\mathbf{k}}=(e^{\beta (\epsilon_{\mathbf{k}}-\mu)}-1)^{-1}$.
Even after using the delta-functions to eliminate integration variables, this is a five-dimensional integral.  To more efficiently evaluate it, we change to center-of-mass and relative coordinates, writing
\begin{align}\label{delta_momentum_K}
\delta&(\mathbf{k}+\mathbf{k'}-\mathbf{k''}-\mathbf{k'''}-\mathbf{Q}) =
\int d^2\mathbf{K} \, \delta(\mathbf{k}+\mathbf{k'}-\mathbf{K})\delta(\mathbf{k''}+\mathbf{k'''}-(\mathbf{K}-\mathbf{Q})) \\
\label{delta_energy_E}
\delta&(\epsilon_{\mathbf{k}}+\epsilon_{\mathbf{k'}}-\epsilon_{\mathbf{k''}}-\epsilon_{\mathbf{k'''}})=
\int d E \, \delta (\epsilon_{\mathbf{k}}+\epsilon_{\mathbf{k'}}-E) \delta (\epsilon_{\mathbf{k''}}+\epsilon_{\mathbf{k'''}}-E).
\end{align}
\end{widetext}
Using the identity
\begin{equation}\label{identity_f}
f_{\mathbf{k}}^0f_{\mathbf{k'}}^0(1+f_{\mathbf{k''}}^0)(1+f_{\mathbf{k'''}}^0) = e^{\beta(E-2\mu)} f_{\mathbf{k}}^0f_{\mathbf{k'}}^0 f_{\mathbf{k''}}^0f_{\mathbf{k'''}}^0 
\end{equation}
we arrive at
\begin{align} \label{P_simplified}
&P =  \frac{U^2a^4}{16\hbar}\beta  \int d E \: e^{\beta(E-2\mu)}  w(E) \\
\nonumber
&w(E)= \int d^2\mathbf{K} \: G_{\mathbf{K}}(E) \\ 
&G_{\mathbf{K}}=\frac{1}{(2\pi)^5}\sum_Q \left[ F^{(2)}_{\mathbf{K}}F^{(0)}_{\mathbf{K-Q}}- 2  F^{(1)}_{\mathbf{K}}F^{(1)}_{\mathbf{K-Q}} + F^{(0)}_{\mathbf{K}}F^{(2)}_{\mathbf{K-Q}}\right]\\
\label{F_m_q}
&F^{(m)}_{\mathbf{K}}(E) = \int d^2\mathbf{q} \; (\phi_{\frac{\mathbf{K}}{2}-\mathbf{q}}+\phi_{\frac{\mathbf{K}}{2}+\mathbf{q}})^m  f_{\frac{\mathbf{K}}{2}-\mathbf{q}}^0f_{\frac{\mathbf{K}}{2}+\mathbf{q}}^0  \nonumber \\
&\qquad \times \; \delta (\epsilon_{\frac{\mathbf{K}}{2}-\mathbf{q}}+\epsilon_{\frac{\mathbf{K}}{2}+\mathbf{q}}-E). 
\nonumber
\end{align}
For each density and temperature we tabulate the 1D integral $F_{\bf K}^{(m)}(E)$ on a $\bf K$ and $E$ grid.  The remaining 3D integral in Eq.~(\ref{P_simplified}) is then efficiently calculated as a discrete sum.  We use a sequence of finer grids to verify that our result has converged to the continuum limit.

The chemical potential $\mu$ is fixed for each $n$ and $T$ by numerically solving
\begin{equation} \label{density}
n(\beta,\mu) = \int \frac{d^{2}\mathbf{k}}{(2\pi)^2} f^{0}_\mathbf{k} \;,
\end{equation}
and the current $j$ is expressed as
\begin{eqnarray} \label{current_j}
j &=& - e\int \frac{d^{2}\mathbf{k}}{(2\pi)^2}  \phi_{\mathbf{k}}
\frac{\partial f^0_k}{\partial \epsilon_k} (\mathbf{v}_{\mathbf{
k}})_x  \nonumber \\
&=& \frac{4t^2ae}{\hbar}\beta \int \frac{d^{2}\mathbf{k}}{(2\pi)^2} \sin^2(k_xa) e^{\beta(\epsilon_{\mathbf{k}}-\mu)} (f^{0}_\mathbf{k})^2
\end{eqnarray}

\begin{figure}
    \centering
    \includegraphics[scale=1.]{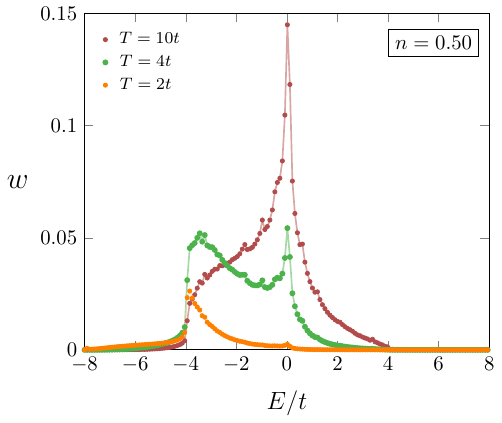}
    \caption{The weight function $w(E)$ from Eq.~(\ref{P_simplified}) as a function of the energy $E$ of the scattering particles.}
    \label{fig:weight}
\end{figure}

One can interpret Eq.~(\ref{P_simplified}) as a Boltzmann sum, where the weight function $w(E)$ corresponds to a density of scattering states.    Figure~\ref{fig:weight} shows $w(E)$ for several temperatures, where $E$ corresponds to the center of mass energy of the scattering particles.  There are three features visible, a sharp peak at $E=0$, a broader peak near $E=-4t$, and a diffuse background running from 
$-8t<E<8t$.  The $E=0$ peak comes from the scattering between two particles who are both near the van-Hove singularity of the square lattice.  The $E=-4 t$ peak is due to the scattering between a low energy particle and one near the van-Hove singularity.  All other scattering events contribute to the background.  As the temperature is lowered, and fewer particles reside near the van-Hove singularity, the relative weight of these features shift.  At low temperature the peak near $E=-4 t$ dominates the calculation of $P$, giving rise to the exponential suppression of both $\rho$ and $\Gamma$.

\section{Low Density Limit}\label{appendix:low_density}

Here we establish that in the resistivity is well-defined in the low density limit $n \to 0$, neither diverging nor vanishing.  We first note that in this limit the chemical potential approaches $-\infty$, and hence the Bose-Einstein distribution is well approximated by the Maxwell-Boltzmann expression 
$f^{0}_{\mathbf{k}} = (e^{\beta(\epsilon_{\mathbf{k}}-\mu)}-1)^{-1} \approx e^{-\beta(\epsilon_{\mathbf{k}}-\mu)}$.  Furthermore,
we can remove the chemical potential dependence by considering the ratio
\begin{eqnarray}\label{f_dilute_ratio}
\frac{f^{0}_{\mathbf{k}}}{n} =  \frac{e^{-\beta\epsilon_{\mathbf{k}}}}{\int \frac{d^{2}\mathbf{k}}{(2\pi)^2} e^{-\beta\epsilon_{\mathbf{k}}} } \;.
\end{eqnarray}
Explicitly making this substitution, we write the resistivity as
\begin{eqnarray}\label{resistivity_dilute}
\rho  =  \frac{\frac{P}{n^4}}{(\frac{j}{n^2})^2}  \;.
\end{eqnarray}
As $n\to 0$, both the numerator and denominator are independent of $n$, and hence $\rho$ approaches a constant.

\section{Variational Procedure}\label{appendix:variational_procedure}

According to \cite{ziman_general_1956,ziman_electrons_2001}, the variational parameter $\xi$ is optimized by maximizing the entropy production from scattering, $T\dot{S}_{coll}=\xi^2P$ subject to the constraint $\xi X = \xi^2P$,  where $X=\mathbf{j}\cdot \mathbf{E}$.  We therefore introduce a Lagrange multiplier $\lambda$, and extremize 
\begin{equation}\label{lagrangian}
\mathcal{L}[\xi] = \xi^2 P -\lambda (\xi^2P-\xi X)\;.
\end{equation}
The conditions $\delta_{\xi} \mathcal{L} =0$ and $\xi X = \xi^2P$ gives $\lambda=2$, which allows us to write the Lagrangian in Onsager's form $\mathcal{L} = -T\dot{S}_{coll} +2T\dot{S}_{field}$ \cite{onsager_reciprocal_1931,onsager_reciprocal_1931-1}, and conclude that
\begin{equation}\label{xi_solution}
\xi  = \frac{X}{P} = \frac{\mathbf{j}\cdot \mathbf{E}}{P}.
\end{equation}


\nocite{*}

\end{document}